**Galileo's Knowledge of Optics and the Functioning of the Telescope ─ Revised**

**Yaakov Zik • Giora Hon**

**Abstract**  What are the historical evidence concerning the turning of the spyglass into an astronomical instrument—the telescope? In *Sidereus Nuncius* and in his private correspondence Galileo tells the reader what he did with the telescope, but he did not disclose the existence of a theory of the instrument. Still, the instruments which Galileo produced are extant and can be studied. With replicas of Galileo's telescopes that magnify 14 and 21 times, we have simulated and analyzed Galileo's practices as he reported them in *Sidereus Nuncius*. On this ground, we propose a new solution to this old problem. We establish the knowledge of optics that Galileo had as it can be read off from the telescopes he constructed and the way he put them to use. Galileo addressed optical difficulties associated with illumination, resolution, field of view, and magnification. His optical knowledge was well thought through, originated as it did in a radically new optical framework.

**1  Introduction**
According to the received view the first spyglass was assembled without any theory of how the instrument magnifies.[1] Between summer 1609 and the end of November 1609, Galileo Galilei (1564–1642), who was the first to use the device as an astronomical instrument, improved its power of magnification up to 21 times and indeed transformed it into a telescope. How did he accomplish this feat? Galileo does not tell us what he did. In a previous publication we proposed a plausibility argument which seeks to show that Galileo could construct a theory of telescope by combining elements of optical knowledge available to him at the time.[2] He could develop it by analogical reasoning based on reflection in mirrors—as they were deployed in surveying instruments—and apply this kind of reasoning to refraction in sets of lenses. Galileo could appeal to this analogy while assuming Della Porta's theory of refraction. He could thus turn the spyglass into a revolutionary astronomical instrument. To be sure, this argument is hypothetical—this is speculative history—but it throws light on how the telescope could have been understood by Galileo, that is, this plausibility argument suggests that Galileo could have a theory of telescope which for obvious reasons he did not divulge.

This plausibility argument diverges substantially from the received view. On the dominating view, the "lightly" theorized empirical methods Galileo had at hand suggests that he would be encouraged to direct spectacle makers and glass smiths to grind lenses of clarity quite beyond what was needed for spectacles at the time, which he would further improve to meet his own optical demands. Galileo was guided—so the traditional argument goes—by experience, more precisely, systematized experience, which was current among northern Italian artisans and men of science. This standard argument underlies the claim that Galileo's practice was an educated extrapolation within spectacle optics. We disagree. We think that Galileo's practice was not an informed extension of the traditional optics of spectacles; rather, the construction and use of the telescope required novel theoretical framework—new optics based on refraction phenomena in system of lenses. We claim that there is no continuity from the optics of spectacles to the optics of telescopes and that Galileo conceived a novel theoretical framework.

---

[1] See, e.g., Van Helden, et al. (2010), Strano (2009), Willach (2008), Molesini, and Greco (1996).
[2] Zik and Hon (2012).

In this paper we respond to the challenge of the received view and establish the knowledge of optics that Galileo had as it can be read off from the telescopes he constructed and the way he put them to use. While our previous work is hypothetical, this paper is factual, based on the extant instruments and Galileo's writings. To be sure, the secondary literature is rich with information regarding the original telescopes of Galileo, and his text on optics; *Sidereus Nuncius* (1610) has been examined and commented upon many times. But we take a fresh look at the problem. We reevaluate the optics of the telescopes attributed to Galileo according to the historical data with advanced optical codes and high standard replicas.[3] With the telescopes at hand, we have simulated and analyzed the practices as Galileo reported them in *Sidereus Nuncius*. We begin by elaborating the received view and exposing its problems (section 2). We analyze the different principles underlying spectacles and telescopes (section 3). The analysis serves as a background for understanding the essential requirements for improving the performance the telescope. We detail our position based on a careful reading of one passage in Galileo's *Sidereus Nuncius* (section 4). We then follow the "optical footprints" in *Sidereus Nuncius* (section 5).

## 2  An exposition of the received view

The received view consists in fact of three different historical conceptions: (1) Galileo's understanding of the spyglass was based on his familiarity with sixteenth century optics, originated in the artisanal tradition, which Ettore Ausonio (c. 1520 – c. 1570), for example, formulated; it consists of a body of practical knowledge of image formation in concave mirrors and convex lenses; (2) Galileo was able to find out the relationship between the focal length of a convex lens and the magnification of the spyglass from a standard method applied by spectacle makers to measure the power of spectacle lenses; and (3) Galileo succeeded in improving the magnification of the spyglass on the basis of his familiarity with workshop practices of spectacle makers which combines elements from (1) and (2).

---

[3] *Elbit Systems, Electro-Optics ELOP Ltd.*, manufactured for us replicas of the lenses of Galileo's telescopes that magnified 14 and 21 times. The lenses made of modern optical glasses are as close as possible to those of Galileo's in terms of their geometrical shape, refractive, and dispersive values (cf., footnote 19 below).

"Conception (1)" rests on the claim that neither Galileo nor any of his contemporaries thought that a concave lens has an optical "length"; *ipso facto* it did not play any role in determining the magnification of the spyglass. The optical performance of a convex lens, as understood in light of the practical optical knowledge from Ausonio to Giovan Battista Della Porta (1535–1615), suggests that it was the diameter of the convex lens which was the critical parameter of magnification.[4] Convex lenses have "length", a term used for denoting the distance at which the lens kindled fire. However, the properties of concave lenses were little understood, since they have no "length" and do not kindle fire. Therefore, convex lenses were the magnifying element and, like in mirror optics, their power of magnification dependent on their diameter.[5]

Galileo's optical practice, so the claim goes in "Conception (1)", is derived from *Theorica speculi concavi sphaerici*, written about 1560 by Ausonio, which Galileo probably saw and copied at the library of Vincenzo Pinelli (1535–1601) in Padua around 1601.[6] The *Theorica* is concerned solely with the optics of reflection; there is no discussion of refraction. It describes how reflected images are seen in a concave spherical mirror. The paths of the incident and reflected rays are traced according to the law of reflection and the place where the object is seen is determined by the cathetus rule. The focal point (*punctum inversionis*) of the mirror in relation to the position of the object can thus be determined.[7] When the observer's eye is placed at the locus of the *punctum inversionis*, the magnified image occupies the complete surface of the mirror. Thus, the magnification is considered a function of the diameter of the mirror, that is, the larger the diameter, the larger the image. The point of combustion, namely, the point at which the concave spherical mirror kindles fire, what Johannes Kepler (1571–1630) named later as the "focus",[8] would be located along the optical axis at a distance equals to half the radius curvature of the interface. The application of the concept of *punctum*

---

[4] Dupré (2005, pp. 171, 173–175).

[5] Dupré (2005, p. 176).

[6] Dupré (2005, pp. 152–160), Favaro (1890–1909, 3: 865–868).

[7] Note that referring to *punctum inversionis* as the focal point of a concave spherical mirror is misleading. In canonical optical texts of the time (e.g., Ptolemy (c. AD 90–c. AD 168), Alhacen (c. 965 – c. 1040), Roger Bacon (c. 1220 –c. 1292), John Pecham (c.1230 – c. 1292), and Witelo (c. 1230 – c. 1280), including Della Porta's *De Refractione*, the geometrical nomenclature used for referring to the point of reflection (*punctus reflexionis*) in mirrors, or the point of refraction (*punctus refractionis*) in refractive elements, was related to the points on the surface where an incident ray is broken. The authors, however, were well familiar with the circumstances in which the image seen in a concave spherical mirror depends upon the relative locations of the eye, the object, and the mirror. Thus the image can be of the same size as, larger than, or smaller than, its object depending on its relative position. The image can also appear reversed or upright, and it can be seen in front or behind the reflecting surface. Della Porta associated the point of inversion (*punctum inversionis*) qualitatively with a locus—not a geometrical point—where the image seen by the eye is turned reverse or upright as a function of the placement of the optical element (i.e., a mirror or a lens) in relation to the observer's eye.

[8] Goldstein and Hon (2005, p. 92).

*inversionis* in magnifying images in curved-mirror optics made—by extension—lens magnification dependent upon the diameters of convex lenses, and not on its "focal length."[9]

In an attempt to shed light on how Galileo's could have realized the correlation between the focal length of a convex lens and the magnification of the spyglass, "Conception (2)" appeals to a standard practice used by spectacle makers to measure the power of lenses. This practice enabled labeling and separating the various iron molds used to grind convex and concave lenses in reference to the age categories of the customers. The power of the lens was measured on a scale, varied from 0.5–15 Venetian punti, which indicated the power of the lenses.[10] To clarify the implications of this practice, the received view turns to a text from 1623 written by Daza de Valdés Benito (1591– c. 1636). This text, related to a practice of a Spanish spectacles maker, is assumed to reflect on the practice used at the second half of the sixteenth century.[11] In this testing procedure the power of lenses, be they convex or concave, could be determined in relation to the customers' age group. Daza de Valdés' test chart is made of two circular figures of different size. The lens under test was placed over one of the figures, the concave lens over the larger figure and the convex lens over the smaller, while the observer looked at them from a distance of about 30 cm. The lens placed over the figure is then lifted away slowly until the figure, seen through the lens, is equal in size to the other figure seen—not superimposed—either with one or both eyes. The distance of the lens from the figure is then measured and compared to a nonlinear scale which referred to lens power range between 2 and 10 units called "grados". Note that this testing method is designed to evaluate the power of singlet lenses, not their focal length. Its accuracy is critically dependent on the viewing distance of the eye from the lens and accurate leveling of the lens.[12]

According to "Conception (3)" Galileo realized that he would need a weak objective lens to "bring" far away objects closer, and a concave lens to sharpen up the image.[13]

> With the procedure of the spectacles makers at hand, ... [Galileo] would have quickly found out, by trying several convex lenses in combination with a standard concave lens, that convex lenses with longer focal lengths resulted in higher magnification.[14]

---

[9] Dupré (2005, p. 171). Note that the reference to "focal length" is anachronistic for Galileo did not have this concept.

[10] Ilardi ( 2007, pp. 226–228), Garzoni (1587, pp. 539–542), Willach (2008, pp. 70–84). Note that spectacle makers do not grind lenses of a certain focal length, but lenses of certain curvature, corresponding to the radius curvature of the grinding molds used. The power of the lens is denoted as a number in terms of its radius curvature or in reference to an age group. For an account of refraction and accommodation related to spectacle lenses, see Fuchs (1901, pp. 661–732).

[11] Dupré (2005, p. 177–178).

[12] Hofstetter (1988, p. 355).

[13] Van Helden (2010, pp. 186–189), Dupré (2005, pp. 174–176).

[14] Dupré (2005, p. 179).

In his search for the right convex lens, Galileo presumably instructed spectacle makers and glass smiths to grind lenses of clarity quite beyond what was needed for spectacles. Galileo then reconfigured and polished the lenses to suit his optical demands and set the most pure lenses in tubes, aligned the lenses, found the right amount of adjustment, and stopped down the aperture.[15] No theory of the instrument was needed, so the argument goes.

### 2.1 *The problem with the received view*

Let us now examine a contemporaneous report on Galileo's telescope. On August 21, 1609, Galileo publicly displayed his newly improved telescope at the Tower of St. Mark to a group of distinguished Venetians. Few days later Galileo showed the instrument to the Signoria and to the Senate. Antonio Priuli, who attended the first presentation, described the instrument Galileo used.[16] The optical properties of this telescope are,[17]

($D_1$) Two lenses: one convex, the other concave.
($D_2$) The diameter of the tube is about 42 mm.
($D_3$) The overall length of the telescope is about 600 mm.
($D_4$) The telescope magnifies nine times.

On the received view, Galileo succeeded in improving the magnification of the spyglass by "systemized experience". The procedure originated in the spectacle makers practices which presupposed the following rules:

($R_1$) Convex lenses determine magnification.
($R_2$) The concave eyepiece has no focal length; it does not play any role in magnification; its role is to sharpen the image.
($R_3$) The standard spectacle lenses available at the time, whether convex or concave, varied between 1.5 diopter to about 5 diopter (i.e., 666 mm and 200 mm respectively).[18]

---

[15] Van Helden (2010, pp. 187–189), Strano (2009, pp. 22–23), Willach (2008, p. 1–3), Dupré (2005, p. 171), Smith, A. M. (2001, pp. 161–162).

[16] Favaro (1890–1909, 19, pp. 587–588): "Che era di banda, fodrato al di fuori di rassa gottonada cremesine, di longhezza tre quarte $^1/_2$ [about 60 cm] incirca et larghezza di uno scudo [about 4.2 cm], con due veri, uno…. cavo, l'altro no, per parte; con il quale, posto a un ochio e serando l'altro, ciasched'uno di noi vide distintamente, oltre Liza Fusina e Marghera, anco Chioza…. E poi da lui presentato in Collegio li 24 del medesimo, moltiplicando la vista con quello 9 volte più…. Presentato in Signoria il giorno d'heri un instrumento, che è un cannon di grossezza d'un scudo d'argento poco più e longhezza di manco d'un braccio [a braccio is about 66 cm], con due veri, l'uno per capo, che presentato all'cchio multiplica la vista nove volte di più dell'ordinario, che non era più stato veduto in Italia."

[17] The keys used in the text are: D – data of Galileo's telescope; R – received view; OR – optical calculations of the received view; OD – optical calculations of the data; efl – effective focal length.

[18] See Van Helden, 1977, pp. 11–12.

(R$_4$) Try several convex lenses in combination with a *standard* concave lens, and find by trial and error that convex lenses with *longer focal lengths* offer higher magnification.

Given R$_1$ – R$_4$ and using *OSLO*,[19] the optical properties of a Galilean telescope that magnifies nine times should be,

(OR$_1$) Standard concave eyepiece of 5 diopter (efl = 200 mm).
(OR$_2$) Convex objective of 0.55 diopter (efl = 1800 mm).
(OR$_3$) The overall length of the telescope for viewing infinite objects is 1.6 meters.

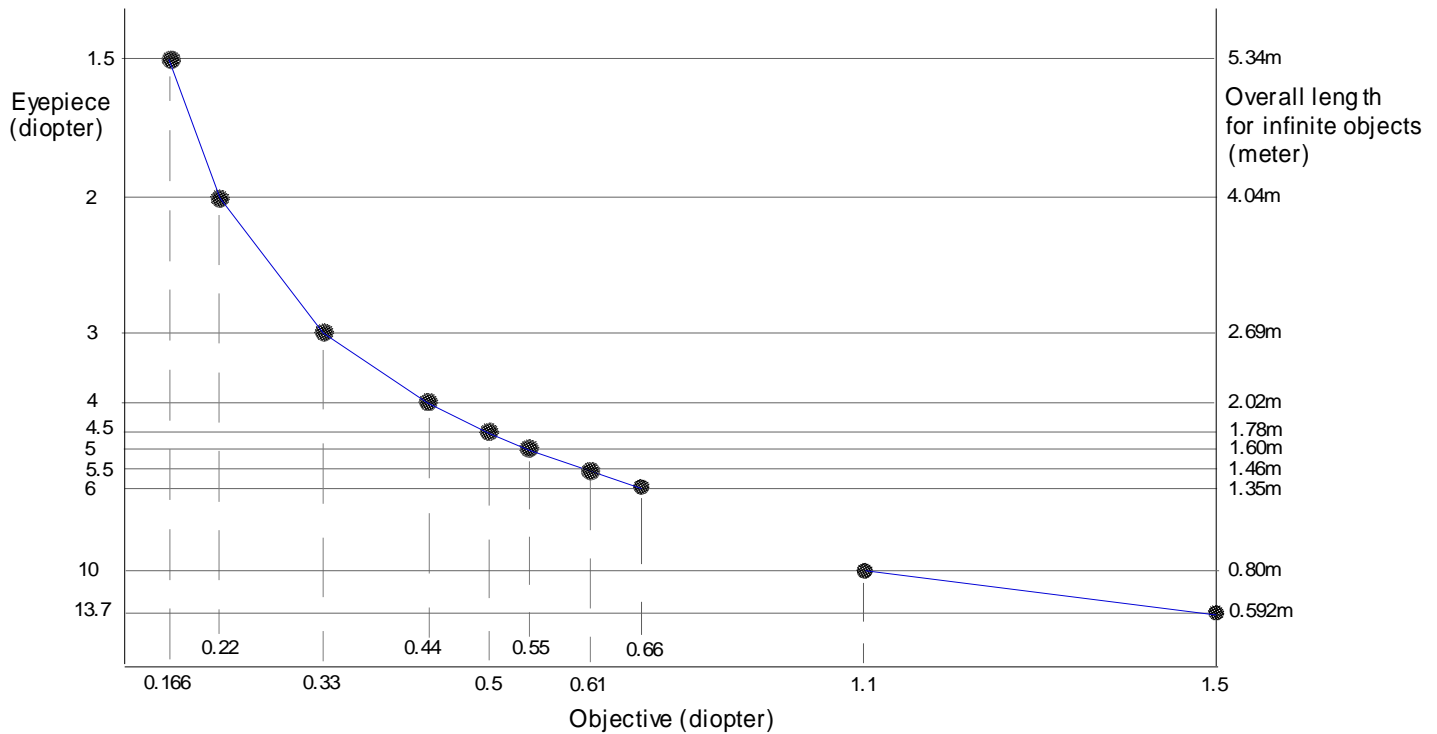

**Figure 1**

The graph in Figure 1 presents the mathematical relations between the optical powers of various lenses combinations (i.e., eyepieces and objectives) producing Galilean telescopes that magnify 9 times. It comprises the following parameters: (1) the abscissa, objective power in diopter; (2) the ordinates (left): eyepiece power in diopter; and (3) the ordinates (right): the overall length of the instrument in meters for viewing infinite objects. Clearly, rather than values OR$_1$ – OR$_3$ one can use different lenses combinations and, of course, one will get different overall lengths. For example, the weaker the eyepiece, the weaker the objective should be, so an eyepiece of 2 diopter (efl = 500 mm) and an objective of 0.22 diopter (efl = 4545 mm) will result in magnification of nine times and overall length of 4.04m. However, by $R_2$ and $R_4$ we are constrained to use only standard concave eyepieces to obtain an instrument that magnifies nine times. All possible combinations one may chose, be they $OR_1$ and $OR_2$ or other optical layouts given in Figure 1, will result in instruments of substantially larger overall length than the length of the telescope which Galileo displayed in Venice.

---

[19] The computations throughout this paper were made with *OSLO* (Optical Software for Layout and Optimization) using the same glass features as those of Galileo's telescope that magnified 21 times, see Greco et al (1993), Zik (1999, pp. 48–51).

We conclude that the telescope Galileo publicly displayed in Venice does not correspond to the telescope which the received view "constructed". No matter what lens combinations one may chose, rules $R_1 - R_4$ imposed a different optical scheme from the one Galileo used to improve the magnification of the telescope he displayed in Venice.

According to data $D_1 - D_4$ and using *OSLO* the optical properties of the telescope Galileo displayed at Venice should be,

   ($OD_1$) Concave eyepiece of 13.7 diopter (efl = 73 mm).
   ($OD_2$) Convex objective of 1.5 diopter (efl = 663 mm).

Note that the focal length of a lens is determined by its power. The magnification of a telescope is determined by the ratio of the power of the eyepiece and the objective and conversely by the ratio of the focal lengths of the objective and the eyepiece. The overall length of a Galilean telescope is determined by the difference of the focal length of its lenses. Given $D_3 - D_4$, only lens combination $OD_1$ and $OD_2$ will comply with the required demand, that is, a significantly stronger eyepiece with a standard (weak) objective. The implications are,

1. Contrary to the received view, an objective of 1.5 diopter is not a significantly *long-focal-length* lens.
2. Contrary to the received view, an eyepiece of 13.7 diopter is not a *standard* concave lens used for sharpening images.
3. The specifications of the objective used by Galileo fit well with the common weak convex lenses available at the spectacle market.

This examination suggests two different optical schemes: (1) Galileo calculated, that is, he had a theory by which the properties of the lenses could have been determined; and (2) Galileo did not have a theory; he simply kept trying more and more powerful eyepieces in the hope that eventually one will work.

Suppose Galileo "kept trying"; it is so to speak a long way from 5 diopter to 13.7 diopter to obtain a lens combination that magnifies 9 times. What did keep his confidence that there will be a power that will eventually work? And recall that he was going, as it were, in the opposite direction. The craftsmen in the market would have advised him to go for a weaker objective with a standard eyepiece, for the eyepiece is just for sharpening the image created by the objective lens as the received view has it. Galileo's practice must have been seen at the time totally counter-intuitive, against the systematized experience which was current among northern Italian artisans and men of science.

We submit that the received view does not help clarifying Galileo's move from the spyglass to the telescope. In fact, we believe that the claim that Galileo pursued a method of "systematized experience", extending the optics of spectacles beyond its standard practices is misleading. A new optics was required in order to turn the spyglass from a toy into an astronomical instrument, an optics which made use of refraction phenomena in a system of lenses of different optical power. In order to consolidate our claim we examine closely the evidence and seek to establish Galileo's optical knowledge as it can be gleaned from his practice.

## 3  The different principles underlying spectacles and telescope

The scale in Figure 2 presents the mathematical relations between the optical power of a lens in diopters and its corresponding focal length in centimeters applied in the measurement of spectacle lenses. The figures underneath the scale are faithful representations of the lens' shapes in relation to their optical power on the scale. At the beginning of the seventeenth century one who sought remedy for poor eyesight had to buy spectacles in the market where the most suitable spectacle lenses were selected from readymade stocks.[20] In the upper part of the figure, we see that the opening through which the light cone enters into the eye is much smaller than the entire diameter of the spectacle lenses. The entering light is limited by the pupil which contracts and expands over a range from about 2 mm in bright light to roughly 7 mm in darkness. In fact, the pupil determines the feasibility of the use of spectacles. The poor quality of optical glass and primitive configuring and polishing techniques resulted in low quality lenses which could hardly serve as visual aids. However, due to the small diameter of the pupil it was still possible to find lenses with reasonable optical performance, especially at the very small sector through which the passing rays were not obstructed by the pupil.

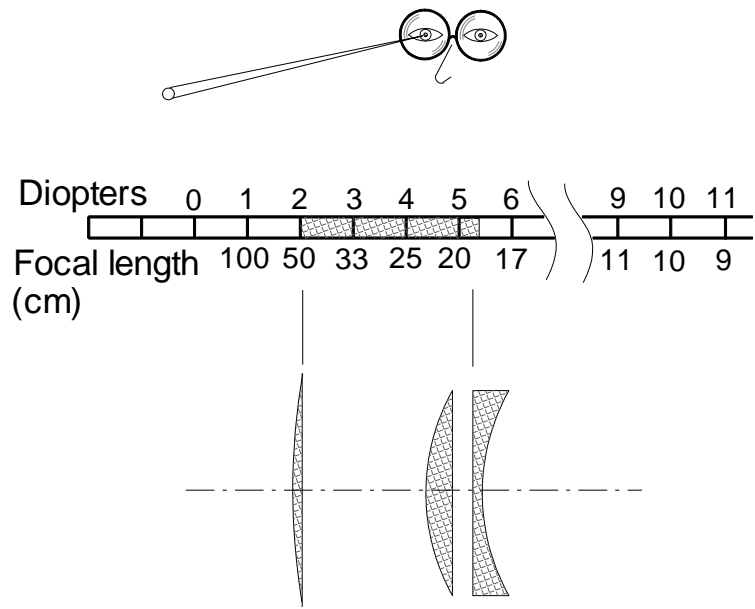

**Figure 2**

---

[20] Ilardi (2007, pp. 226, 230–235).

In general, any combination of convex and concave spectacle lenses results in a spyglass that can magnify up to 3 times. Due to the low power of magnification, even lenses of poor quality are sufficient for the job. Playing around with pairs of spectacle lenses led several Dutch lens makers to appeal for a patent on the spyglass in October, 1608.[21] This in effect is the whole story of the invention of the Dutch spyglass.

The construction of an astronomical telescope is more than just a mechanical procedure of mounting lenses made of sufficiently pure glass in tubes and the application of a diaphragm to its objective lens. Intricate relations exist among a set of different parameters: focal lengths of lenses, overall lens length of the telescope, diameters of entrance and exit pupils, diameter of the observer's pupil, and field of view of the telescope. Optimization of the performance of a telescope requires testing method and controlling the interplay among material and optical features.

A refractor telescope requires a combination of at least two lenses of different optical powers. Figure 3 presents the properties of telescope lenses in comparison with spectacle lenses (placed in the rectangle). In the upper part of Figure 3 the telescope gathers light through the whole diameter of the entrance pupil located at the objective plane. Then, the light travels further through the eyepiece on its way to the eye.

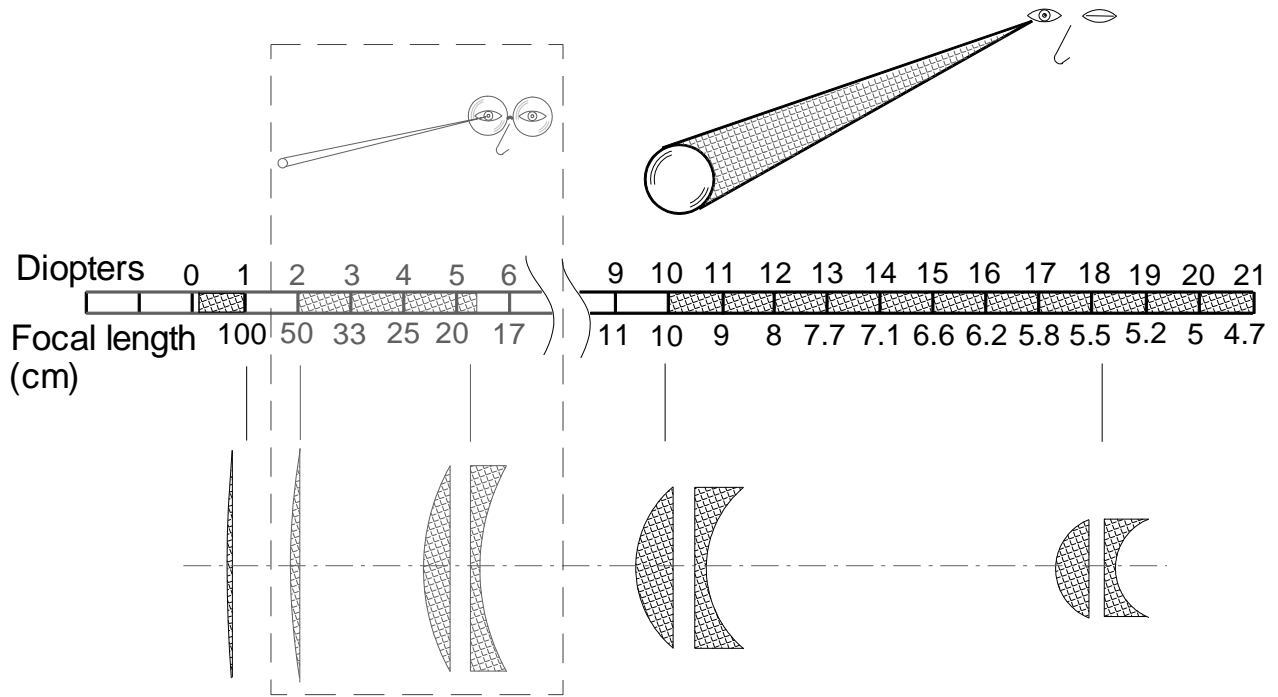

**Figure 3**

Imperfections of the optical elements, especially when the magnification is getting higher, constitute major obstacles to the performance of the telescope. While the optical power of the objective does not exceed 1 diopter, the optical power of the eyepiece, like in the telescopes attributed to Galileo, varied between 10 to 21 diopter (efl = 10 cm and about 5 cm respectively). The objective focal length of a telescope is significantly long and the focal length of the eyepiece is significantly short in comparison with the properties of lenses needed for spectacle lenses. The radius curvature of the objective is much larger and the radius curvature of the eyepiece is much smaller than any of the radii needed for spectacle lenses, as can be seen in Figure 3. The imperfections could be compensated by either choosing the purest glass from the materials available, or using concave eyepieces which are less thicker at the centre in comparison with the convex lenses.

---

[21] Van Helden (2009, pp. 3–8; 1977, pp. 20–25).

    To clarify this point let us examine an interferogram of the concave surface of the eyepiece in Galileo's telescope that magnifies 14 times.[22] The analysis in Figure 4 is made with an optical code which renders a synthetic grayscale plot of the interferogram and a map of the spherical wave front emerging from the exit pupil of the system. In addition we show the charts of MTF curve (modulation transfer function), PSF (point spread function) map and PSF surface.[23] The analysis presented at the left side of Figure 4, shows the reflection pattern of the whole diameter of the eyepiece aperture stop (11 mm). The highly distorted synthetic plot and the wave front map fit well with the pattern of the MTF curve which is indicative of values that cannot be resolved in details by the observer's eye. The resulted large blur spot masking the image plane is shown in the charts of the PSF map and PSF surface. A Galilean telescope has a small exit pupil located inside the telescope. The circle drawn at the center of the interferogram at the right side of Figure 4 is equal in diameter to the diameter of the exit pupil of the telescope (1.836 mm). The reflection pattern of a much smaller sector of the eyepiece, in comparison with the former analysis, results in a significantly less distorted synthetic plot and wave front map. The pattern of MTF curve is indicative of a lens that performs within the diffraction limit. The PSF map and the PSF surface charts attest to a sharp undistorted image. In the lower part of Figure 4 the circle representing the exit pupil of the telescope is moved off-axis towards the left upper rim of the eyepiece. The reflection pattern of the lens results in noticeable distortions in the synthetic plot and wave front map.[24] The pattern of MTF curve is indicative of a lens with poor off-axis optical performance. The PSF map and the PSF surface charts attest to the severe degradation of image quality.

---

[22] Greco et al (1992, p. 101, interferogram *b*). The interferometric images of Galileo's telescopes are Copyright of *Nature* and *Applied Optics*. We thank these Journals for giving us the permission to reproduce these images.

[23] For an account of image evaluation, see Smith, W. (1990, pp. 327–361), Malacara (1992, pp. 77–86).

[24] The wave front error in this interferogram corresponds to RMS OPD (Root Mean Square Optical Path Difference) of 0.2 wavelengths in comparison with RMS OPD of 0.04 wavelengths resulted in the former interferogram. A diffraction limited system is considered to have RMS OPD values better than 0.07 wavelengths.

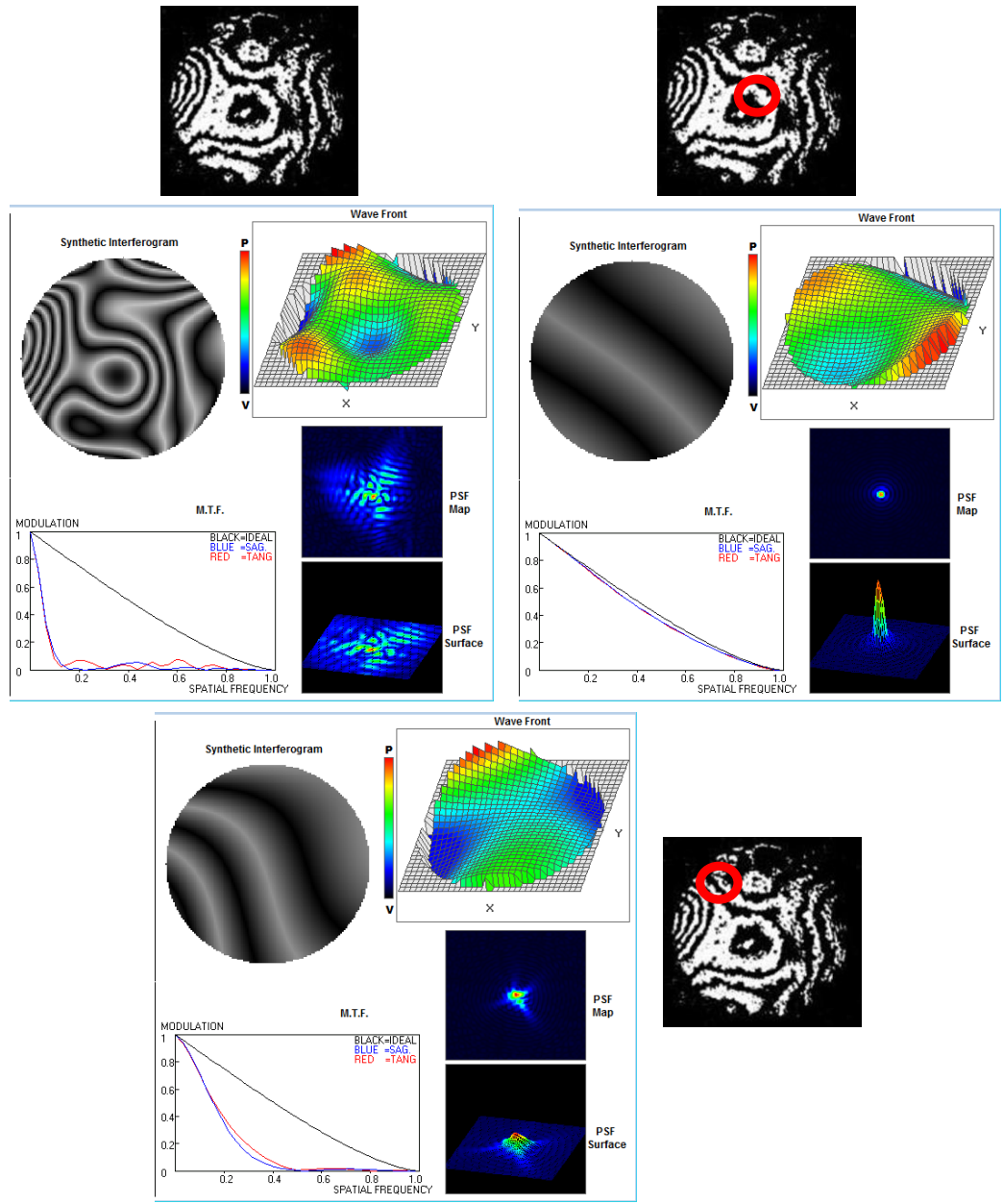

**Figure 4**

Galileo's 14X and 21X telescopes are typical optical systems in which the entrance pupil coincides with the objective. Thus with an entrance pupil of 16 mm and exit pupil of 0.77 mm, in the 21X Galilean telescope (1) the field of view is limited by the diameter of observer's pupil; (2) the axial bundle of rays are passing through a central thin cross section of much smaller diameter of the concave eyepiece; and (3) for paraxial (i.e., the region about the optical axis) image points, the throughput of the telescope is less liable to interferences caused by the poor quality of the glass and inadequate configuring and polishing techniques of lenses at the time.

## 4  The alternative claim

We begin by revisiting Galileo's *Sidereus Nuncius*. Galileo informed the reader what properties a good astronomical telescope should have. We concentrate on the following passage which appears in the first section of this treatise:

> For it is necessary first that ... [the observers] prepare a most accurate glass that shows objects brightly, distinctly, and not veiled by any obscurity, and second that it multiply ... [the observed objects] at least four hundred times and show them twenty times closer.[25]

Galileo singles out four features:

1. The objects should be seen bright [*pellucida*].
2. The objects should be seen distinct [*distincta*].
3. The objects should not be veiled by any obscurity [*nulla caligine obducta*], and
4. The objects should be seen at least twenty times closer [*bisdecuplo viciniora commonstrabit*].

We appeal to modern terminology and, corresponding to these four features, remark:

1. The telescope should not filter the illumination of the objects below certain degree when the illumination-contrast in the image is less than the smallest amount that the eye can detect.
2. The telescope should have sufficient resolution so that fine details of the objects can be resolved by the eye.
3. The objects should not be seen blurred due to inappropriate focusing of the instrument.
4. The magnification should be at least 20.

These four features determine the operational limits which Galileo set for his astronomical telescope. In addition to effects arising from poor illumination and insufficient resolution, the blurred image could be the result of inappropriate focusing. Galileo knew this fact since he looked at a resolution target while calibrating the instrument and measuring its power of magnification (see section 5.2, below).

   The telescope is a complex visual system. Its complexity arises from the combination of different optical properties associated with each of the elements which comprise the instrument (e.g., lenses, tubes, lenses holders, and aperture stops). For the instrument to function properly all these optical elements must be adjusted optimally. This is obtained by testing, focus adjustment, and fine calibration—a tradeoff process among the elements.

   Given that Galileo succeeded in transforming a toy into an astronomical instrument, a set of five questions arises with respect to its application:

---

[25] Galileo ([1610] 1989, p. 38), Favaro (1890–1909, 3: p. 61): "Primo enim necessarium est, vt sibi Perspicillum parent exactissimum, quod obiecta pellucida, distincta, & nulla caligine obducta repraesentet; eademque ad minus secundum quatercentuplam rationem multiplicet; tunc enim illa bisdecuplo viciniora commonstrabit."

1. Did Galileo adjust the focus of the telescope?
2. Did Galileo develop techniques for testing and fine tuning of the telescope?
3. Did Galileo take measures to minimize the adverse effects created by optical aberrations?
4. How did Galileo convince his readers to believe in what he had seen with the telescope?
5. Did Galileo know how is the power of magnification related to the magnitudes of the object and the image seen by the eye?

Answers to this set of questions will allow us to determine what was Galileo's knowledge of optics?

Galileo did not disclose his theory of the telescope, nor did he explain how he produced an improved instrument.[26] However, Galileo did report on his usage of the instrument. In *Sidereus Nuncius* and in his private correspondence Galileo told the reader what he did with the telescope. Two telescopes and an objective lens attributed to Galileo are preserved in the Museum Galileo in Florence.[27] We have studied Galileo's telescopes and examined closely the practices he reported in *Sidereus Nuncius*. These are our sources.

## 5 Galileo's optical knowledge
### 5.1 *Did Galileo adjust the focus of the telescope?*
In a letter written on January 7, 1610, Galileo remarked,

> It would be well if the tube [*cannone*] could be capable of being elongated or shortened a little, about 3 or 4 fingers (*dita*), because I have found that in order to see objects close by distinctly, the tube must be longer, and for distant objects shorter.[28]

Galileo explained that adjustment of the tube is essential for obtaining sufficient distinctness of the observed objects. This adjustment yields focus shift. The process of focusing a telescope has two goals. First to determine the distance between the lenses at which the objects are seen most distinctly; second, to compensate for the observer's individual eyesight condition in terms of near or far sightedness. Focal adjustment involves careful design and production of tubes and lens holders so that they could properly accommodate the lenses.

To have better understanding of the procedure involved in mounting and focal adjustment, we examine the Galilean telescopes that magnifies 14 times and 21 times

---

[26] On the tension between secrecy and transparency in Galileo's *Sidereus Nuncius*, see Biagioli (2006, pp. 14–19, 77–134), Zik and Van Helden (2003).

[27] Greco et al (1993), Zik (1999, pp. 48–49).

[28] Favaro (1890–1909, 10: p. 278) letter 259, our translation. The letter was probably addressed to Antonio de Medici. *Dito* is about 1.85 cm; the distance between the lenses could change then up to 7.4 cm, see Biringuccio ([1540] 1943, p. 457).

(Figure 5).[29] The length of the main tube of the telescope that magnifies 14 times (Figure 5a) is 1197 mm. The objective holder is fixed to an adjustable objective tube of about 152 mm in length which can be moved inside the main tube. The eyepiece holder is fixed to an adjustable eyepiece tube of about 192 mm in length which can be moved inside the main tube. The mounting of the lenses holders allows a minimal distance of 1228 mm between the objective and eyepiece and a maximal distance of 1342 mm. The overall length of the telescope is determined by the difference between the focal length of the objective and eyepiece is, 1236 mm. Our computation of the optimal distance between the objective and eyepiece for an infinite object is 1234.7 mm. The length of the main tube of the telescope that magnifies 21 times (Figure 5b) is 837 mm. The mounting of the lenses holders, attached to the main tube, allow a minimal distance of about 920 mm between the objective and eyepiece and a maximal distance of about 938 mm. The overall length of the telescope is 932.5 mm. Our computation of the optimal distance between the objective and eyepiece for an infinite object for this telescope is 930.72 mm.

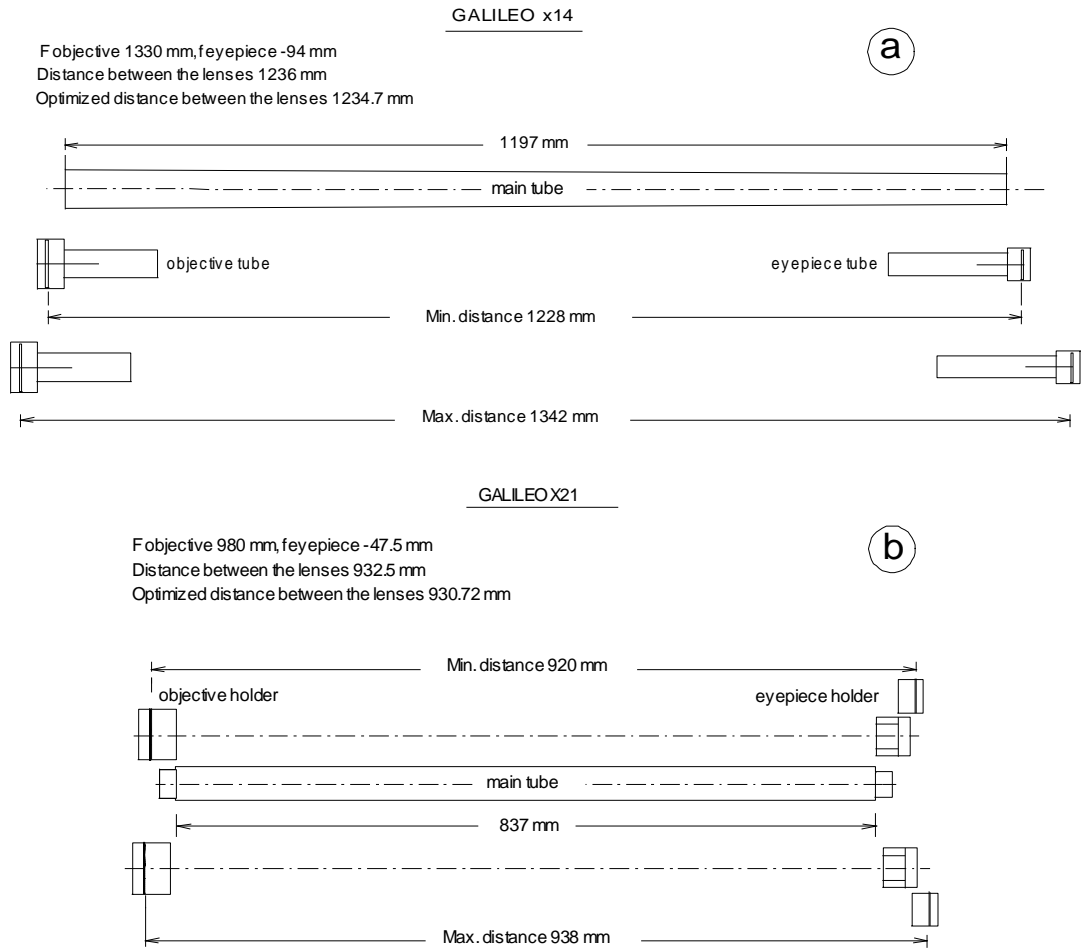

**Figure 5**

---

[29] For Galileo's telescopes that magnify 14 and 21 times, see Van Helden (1999, pp. 30–32), Baxandall (1924, pp. 141–142), Pettit (1939).

Galileo's telescope that magnifies 14 times yields a focus shift of 70.7 mm for objects ranges from 25 m to infinity, while the maximal focal travel available in this telescope is about 114 mm. Galileo's telescope that magnifies 21 times yields a focus shift of 38.4 mm for objects ranges from 25 m to infinity. An examination of the actual telescope which magnifies 21 times reveals that the maximal focus travel available in this telescope is about 18 mm. We ask, What does this discrepancy between focus shift of 38.4 mm and one of 18 mm indicate? Galileo's telescopes consisted of a main tube and lens holders which enabled certain limited focal adjustments, rather than two relatively long tubes with lenses placed at their two opposite ends which allowed much larger amount of focal adjustments. One possibility is that Galileo calculated in advance the focus shift required by the telescope for optimal observation of infinite objects, as the design of this telescope suggests. In other words, Galileo probably did not aim at the full focus shift possible and restricted the range in advance by appealing to novel calculation of which he left no trace. We argued elsewhere that a method of computation could have facilitated Galileo's design process.[30] Be that as it may, the question persists, How did Galileo know in advance the required range of focus shift?

In *Sidereus Nuncius* and in his private correspondence Galileo did not provide any details concerning the dimensions and "lengths" of the lenses he used; nor did he disclose the physical layout of his telescopes or how they were produced. But Galileo's instructions in his letter of January 7, 1610, attests to his understanding of the focus shift needed for adjusting the telescope. Clearly, the layouts and elements of the telescopes attributed to Galileo were not a product of a certain standard production line. Each one of the telescopes was constructed to accommodate lenses of specific length inserted in a different setup of tubes and lens holders. This highly suggests the possibility that Galileo had knowledge of the "lengths" of the lenses he used, and that he could calculate and instruct in advance the length of the tubes and the required focus shift so that the telescope would work optimally.

**5.2** *Did Galileo develop techniques for testing and fine tuning of the telescope?*
Galileo sought convincing arguments to persuade the community that his astronomical observations are reliable. He invented a method for measuring magnification which, at the same time, could test and tune the telescope:

> In order that anyone may, with little trouble, make himself more certain about the magnification of the instrument, let him draw two circles or two squares on paper, one of which is four hundred times larger than the other, which will be the case when the larger diameter is twenty times the length of the other diameter. He will than observe from afar both sheets fixed to the same wall, the smaller one with one eye applied to the glass and the larger one with the other, naked eye. This can easily be done with both eyes open at the same time. Both figures will then appear of the same size if the instrument multiplies objects according to the desired proportions.[31]

---

[30] Zik and Hon (2012, pp. 459–460).
[31] Galileo ([1610] 1989, p. 38).

We can immediately see that Galileo intended this novel method of measuring magnification to work also as a calibrating procedure. The simultaneous comparison of two geometrical figures of known magnitude—one seen through the telescope and the other, at the same time, by the naked eye as a resolution target—constitutes an ingenious method for computing magnification. Here we have an independent method for calculating magnification which in turn expresses a relation between the magnitude of the object and the magnitude of the image produced by the telescope. The importance of the method cannot be exaggerated; Galileo invented a method for measuring magnification which, in turn, expresses another relation, that is, a relation between the object and its image.

While looking through the telescope, Galileo does not report on several things that surely could not have escaped his eyes. If one were to examine the same target that Galileo described, in the same way he did it, one would realize that: (1) clear and distinct image of the target is obtained when the telescope is properly focused; and (2) the lenses—which were not corrected for aberrations—may introduce certain degree of disturbances into the image. We emphasize that this two important aspects are concomitant to measuring the magnification of the telescope. In sum, Galileo set empirical criteria for testing and tuning the telescope so that the goal of a clear and distinct image would be obtained.

### 5.3 *Did Galileo take measures to minimize the adverse effects created by optical aberrations?*

Galileo realized that stopping down of the aperture of the telescope is essential for further improvement of the distinctness of the observed objects. He remarked,

> It would be better if the convex lens, which is the furthest from the eye, were in part covered, and that the opening which is left uncovered be of an *oval* shape, because in this manner it would be possible to see objects much more distinctly.[32]

Galileo's advice reveals his awareness that stopping down and even modifying the shape of the aperture (i.e., controlling the light gathering of the instrument) would be an effective measure for improving the distinctness of the image.

The Galilean telescope is made of singlet lenses; it suffers from axial and lateral chromatic aberration, especially with higher magnification, which produces colored images on the edge of the field of view. In *Sidereus Nuncius* Galileo reported how he tried to stop down the aperture of his telescopes with diaphragms pierced with small holes. It is most likely that in the process of testing his telescopes Galileo noticed the tradeoff effects on the brightness and distinctness of the image, caused by narrowing the aperture. The appearance of stars and planets was affected by accidental brightness and sparkling rays reflected upon the pupil of the eye which enlarged the apparent diameters of celestial objects.[33] By changing the aperture stop the brightness (i.e., illuminating irradiance) at the image was also altered. Galileo remarked that the way to improve the blurred appearances of the bright objects is twofold: (1) by applying a diaphragm, which would function like the pupil of the eye in cutting off the adventitious and accidental rays encircling the bright objects; and (2) by improving the magnification of the telescope so that the diameter of the bright objects will match the diaphragm diameter and the adventitious rays would not pass through the telescope.[34] In such a telescope the adventitious and accidental rays are removed from the bright celestial bodies by design. At the same time, by this very design, the images of the celestial globes are enlarged; the globes appear increased in size by a much smaller ratio in comparison with their blurred appearance.[35]

---

[32] Favaro (1890–1909, 10: p. 278) letter 259, our translation, italics added.
[33] Galileo ([1610] 1989, pp. 57–59), Zik (2002, pp. 460–464).

We know that the telescopes Galileo built had small apertures with respect to their overall "length". Consequently, the Seidel aberrations in monochromatic light in general and their effects on paraxial image points in particular are not critical once the appropriate focal adjustments have been taken.[36] Under optimal conditions, it is not necessary that all aberrations be completely removed, but only that they be reduced to tolerable proportions. Accordingly, focal adjustments and modifying the diameter of the aperture stop proved to be an effective means of controlling most of the aberrations in telescopes composed of singlet lenses.

To demonstrate the efficiency of the aperture stop in reducing the aberrations of an actual Galilean telescope we analyze a Ronchigram of the whole diameter (i.e., 37 mm) of the objective lens of Galileo's telescope that magnifies 21 times (Figure 6).[37] The synthetic plot at the upper left side of Figure 6 is a fair representation of the transmission pattern which could have been rendered by a digital phase shift interferometer for the whole diameter of this lens. In accordance with Ronchi's analysis of this lens, the synthetic plot demonstrates combined patterns resulted from relatively high values of low order astigmatism, coma, and spherical aberration. The white circle on the synthetic plot denotes a 16 mm aperture stop of this telescope adjusted to the less distorted sector of the synthetic plot. The resulted fringe pattern of the stopped sector shown at the right upper side is markedly similar to the real transmission patterns of this objective lens (stopped down at 16 mm), shown at the lower right side of Figure 6.[38] The analysis of the real transmission patterns is presented at the left lower side of Figure 6. The light passing through the aperture of the objective renders a less distorted synthetic plot and wave front map. The presence of low order astigmatism, coma, and spherical aberration is significantly reduced.

---

[34] Galileo ([1610] 1989, p. 58).

[35] For example, the Moon is an object that presents a larger image than the field of view of the telescope. It therefore leaves no room for the adventitious rays to reach the eye, see Brown (1985, pp. 491–492).

[36] Seidel aberrations of a lens system are spherical aberration, coma, astigmatism, Petzval curvature, and distortion.

[37] The middle Ronchigram presented in Figure 3, in Ronchi (1923, pp. 801–802). For a discussion of Ronchi Test, see Malacara (1992, pp. 321–365).

[38] Greco et al (1993, p. 6222, Figure 7, interferogram *c*).

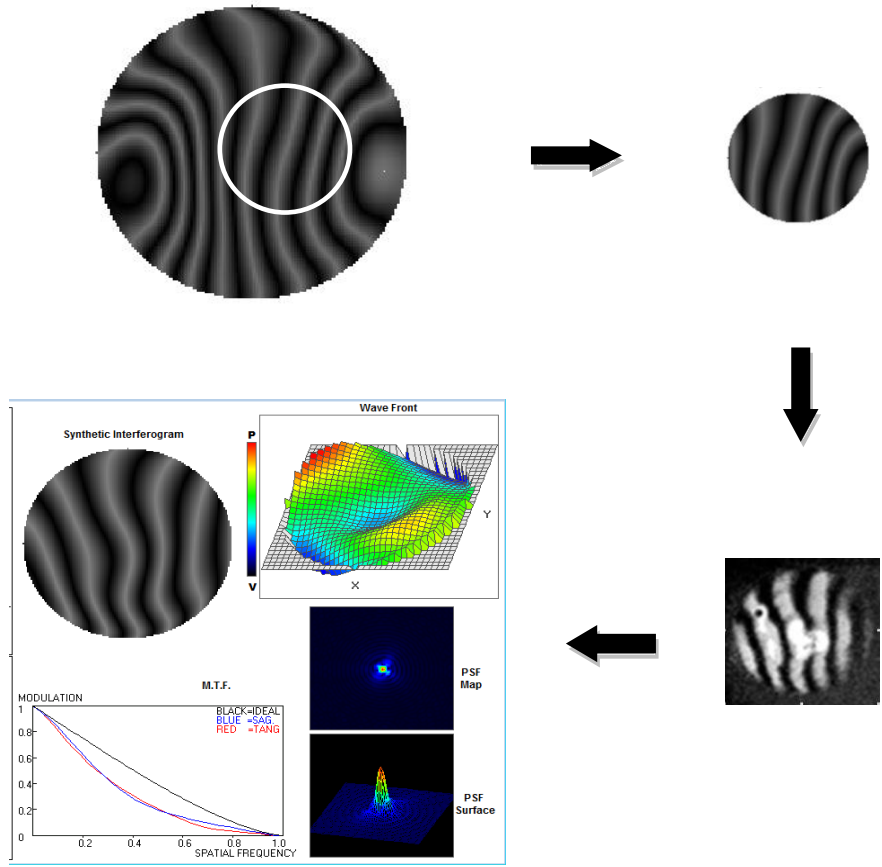

**Figure 6**

The aperture diameter of the telescope that magnified 14 times was 26 mm, while the telescope that magnified 21 times was stopped at 16 mm. Galileo controlled the effects of the accidental brightness and sparkling rays using apertures of various diameters and even slight modifications of their circular shape to ease the effects caused by either irregularities of the surfaces of the objective lens (i.e., small values of low order astigmatism) or inhomogeneous sector of the objective lens.[39]

Indeed, until the end of 1610 only telescopes made by Galileo were equipped with aperture stops. This is evident from a letter sent by Christopher Clavius (1538 – 1612) to Galileo in December 17, 1610, in which he inquired,

> We have seen here in Rome some telescopes which you [Galileo] have sent, which had very large convex lenses covered so that a very small opening is left over. We would like to know what is the purpose for using such large lenses if they were 194> partly covered? Some of us think that these lenses are made so large, so that the entire opening may be exposed at night, in order to better see the stars.[40]

---

[39] Favaro (1890–1909, 10: pp. 485, 501–502), Zik (1999, pp. 58–59), Galileo ([1610] 1989, p. 14).

The Jesuit mathematicians of the Collegio Romano heard about Galileo's discoveries even before *Sidereus Nuncius* was published. In an effort to observe and verify Galileo's discoveries they used their own constructed telescope, but with little success.[41] It is not surprising that Clavius approached Galileo with such a relevant question about the aperture stop. Galileo's answer would satisfy a scholar such as Clavius. Galileo referred to the grinding techniques of the lenses and then to the enlargement of the diameter of the objective lens that, by taking the cover off, increases the field of vision. Nevertheless, adhering to his secrets Galileo did not mention a fact he knew very well, namely, the vital role of the aperture stop in improving the quality of the image.[42]

### 5.4 *How did Galileo convince his readers to believe in what he had seen with the telescope?*

It has been argued that at the outset of *Sidereus Nuncius* Galileo undertook to explain how his new telescope works.[43] But Galileo's account was aimed at achieving different purpose. Galileo described what an observer sees through the telescope and elaborated how the instrument could be used for measuring angular distance between celestial objects (Figure 7):

> After such an instrument has been prepared, the method of measuring distances is to be investigated, which is achieved by the following procedure. For the sake of easy comprehension, Let ABCD be the tube and E the eye of the observer. When there are no glasses in the tube, the rays proceed to the object FG along the straight lines ECF and EDG, but with the glasses put in they proceed along the refracted lines ECH and EDI. They are indeed squeezed together and where before, free, they were directed to the object FG, now they only grasp the part HI. Then, having found the ratio of the distance EH to the line HI, the size of the angle subtended at the eye by the object HI is found from the table of sines (*tabulam sinuum*), and we will find this angle to contain only some minutes, and if over the glass CD we fit plates perforated some with larger and some with smaller holes, putting now this plate and now that one over it as needed, we form at will angles subtended more or fewer minutes. By this means we can conveniently measure the spaces between stars separated from each other by several minutes with an error of less than one or two minutes.[44]

---

[40] Favaro (1890–1909, 10: 485).
[41] Biagioli, (2006, pp. 86–97), Lattis (1994, pp. 180–187).
[42] Galileo's letter of December 30, 1610, see Favaro (1890–1909, 10: 501–502).
[43] Strano (2009, pp. 20–21), Dupré (2005, pp. 174–175), Smith, A. M. (2001, pp. 149–150).
[44] Galileo ([1610] 1989, pp. 38–39).

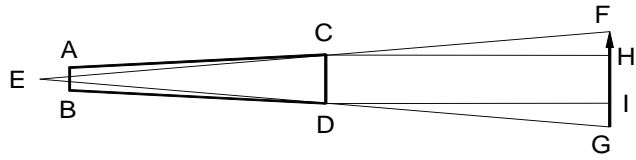

**Figure 7**

Though the account is made in geometrical terms nothing here explains the optical principles of the telescope or the role of the eyepiece. Galileo's did not intend to explain the principle of the telescope, but to present his readers with a method for measuring apparent angular distances between terrestrial objects and heavenly bodies. To clarify his argument Galileo depicted the path of the visual rays seen from the eye as an analytical tool. Thus, the question whether vision occurred due to the extramission or intromission ray propagation did not play any role in Galileo's considerations; in both frameworks, however, the rays travel back and forth along the same optical path.[45] For Galileo the measurement of angular distance was critical for convincing the reader that telescopic astronomical observations are reliable. This quantitative method was applied to calculate the angular magnitude of either the resolution targets or an object FG of known magnitude and distance. With the objective lens mounted in the tube, the refracted rays CH and DI encapsulate only the segment HI of object FG. The segment HI thus appears commensurately larger in relation to object FG. The ratio between segment HI to the distance EH from the observer's eye, enables one to compute an apparent visual angle (i.e., the apparent angular diameter of the entrance pupil as seen by the eye).[46] In effect, this angle equals to the apparent field of view of the telescope. For example, if an object occupy's only one third of the diameter of the entrance pupil seen by the eye, the angle subtended at the eye would be one third of the apparent field of view.

    We appeal to modern terminology and make the following remarks. While testing his telescopes Galileo must have noticed the differences between day and night fields of view of the telescope he used. The overall lens length of Galileo's telescope that magnifies 21 times, focused at infinite object, is 934.5 mm. Given that at least 50% of the illumination passes through the optical system and an eye relief of 10 mm, the apparent field of view of this telescope in day light for an eye pupil diameter of 3 mm is about 10'. At night when the diameter of the eye pupil expands to 7 mm, the apparent field of view of the telescope is increased to about 20'.[47] The evaluation of the distances between stars in reference to a field of view of about 20' was a tricky procedure. To improve the accuracy of his measurements Galileo tried—as the above quotation indicates—to fit plates perforated with holes of different diameters to the objective lens.

---

[45] Smith, A. M. (2001, p. 157; 2010, pp. cii–ciii).

[46] The path EH is not a straight line; it is the refracted line ECH. Since only small angles are involved it is a reasonable approximation on the part of Galileo to consider the path a straight line.

[47] Note that in reference to the faculties of vision Della Porta explained how the field of view expands in the dark because of the dilation of the pupil, see *De Refractione* (1593, 3: pp. 74–76; 77–80).

Galileo defined the apparent field of view of the telescope as the angle *a* subtended at the eye, E, by JK (Figure 8), that is, the angle subtended by segment HI of the object FG (Figure 7). The claim is that reducing the field angle of the telescope would make the comparison between that angle and the observed angular distance of two stars more easily obtained.

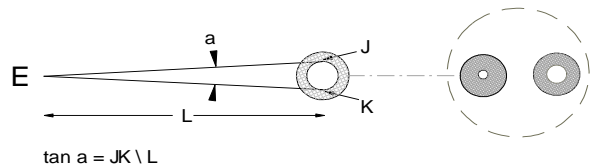

tan a = JK \ L

**Figure 8**

Galileo considered the apparent diameter of Jupiter as equal to a unit of one minute.[48] In *Sidereus Nuncius* Galileo reported the distances among Jupiter and its four Moons in units of minutes and fractions of minutes he denoted in seconds. This method, according to Galileo, yielded an observational error of less than one or two minutes (see the above citation). The practice, however, was not based on precise measurements taken in reference to a defined scale, but on an estimation of how many Jupiter's diameters filled up the distance between the planet and the measured Moons.[49] Galileo was probably aware of the shortcomings of this method for he came up with a much better method for measuring angular distance (see section 5.5, below).

In sum, Galileo developed a method for measuring the angular distances between objects in reference to the apparent field of view of the telescope. This is a novel method of analysis and demonstration; Galileo applied quantitative optical nomenclature, alien to the perspectivist qualitative analysis of visual perception. Galileo exploited this knowledge for the measurement of angular distances between Jupiter and its four Moons.[50] Galileo's account is stated in geometrical terms. He could thus convince his readers to believe in his astronomical discoveries he made with the telescope and yet disclosing none of his secrets.[51]

### 5.5 *Did Galileo know how the power of magnification is related to the magnitudes of the object and the image seen by the eye?*

Galileo measured the angular distances between planets and stars in relation to the apparent field of view of the telescope he used. On January 31, 1612, he noted in his observational records that for the first time he applied a new and most accurate device for measuring the intervals between Jupiter and its Moons.[52] Recall that the telescope magnifies the apparent angular size of a distant object, angle *a* (Figure 9) by presenting to the eye an image which subtends a larger angle than *a*, that is, angle *b′*. The magnification is defined by the ratio, $M = b′ / a$. When one looks through the telescope at, say, Jupiter and one of its Moons, the image, Im, is seen under angle *b′* which is determined by the magnifying power of the telescope. Now, when one looks at the same time with the other (unaided) eye on Jupiter and one of its Moons through a defined grid (reticulum) placed at distance EO from the eye along the telescope's tube, the images presented by the telescope and the grid are seen superimposed.[53] The image, Im, could be measured by the superimposed grid and its angular magnitude calculated, that is, the angular magnitude in the image space is the ratio, $Di = Im / EO$. The angular magnitude of the object in the object space is, $Do = Di / M$. In this way Galileo significantly improved his measuring techniques and calculated accurately the angular distances between stars and planets, separated from each other by short intervals.

---

[48] Drake (1983, pp. 217–219).

[49] For an analysis of the accuracy of Galileo's observations in *Sidereus Nuncius*, see Bernieri (2012).

[50] Galileo ([1610] 1989, p. 66). On Galileo's diagrams in *Sidereus Nuncius* see, Drake (1983, pp. 212–216), Favaro (1890–1909, 3: pp. 427–428), Gingerich and Van Helden (2003).

[51] Smith A. M. (2001, pp. 156–157; 1981).

Galileo's method discloses his knowledge that the relation between the magnitude of the image (in the image space) and that of the object (in the object space) is correlated by the magnification of the telescope. Indeed, he drew consequences from the definition of magnification.[54] The principle of optical magnification which Galileo had conceived in summer 1609 underpins the working of the new micrometric device he applied to measure accurately magnitudes of celestial bodies.

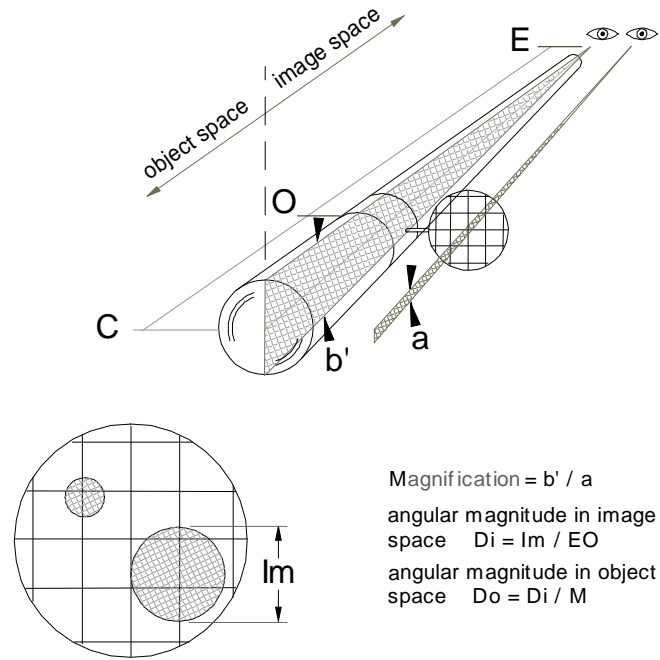

**Figure 9**

---

[52] Favaro (1890–1909, 3: p. 446).
[53] Zik (2001, pp. 267–269), Drake and Kowal (1980, pp. 54–55, 57), Borellio (1666, pp. 143–144, 208 Tavola V).
[54] Zik and Hon (2012, pp. 456–458).

This account shows that Galileo mastered the basic principles of the telescope, that is, a telescope magnifies distant objects by presenting them to the eye in a greater angle than the angle of the objects subtended by the naked eye.[55]

## 6 Conclusion

Careful reading of Galileo's writings clarifies that he ingeniously combined two distinct bodies of optical knowledge which together offer insight into the working of the telescope. Firstly, in his letter of August 24, 1609 to Leonardo Donato, and the letter of August 29, 1609 to Benedeto Landucci, Galileo wrote that his new instrument—that magnified 9 times—was constructed on the basis of knowledge derived from the most recondite speculations of *perspective*.[56] This body of knowledge corresponds, *inter alia*, to the manipulation of visual angles and linear magnitudes (i.e., distances, heights, widths, and breadths) of visible objects. By applying the rules of measurement by sighting, using an optical device such as his military compass in its function as a quadrant, Galileo measured the apparent sizes of remote objects. From this body of knowledge Galileo could have learned that the geometry of magnification in surveying instruments is analogous to that of the telescope.[57]

In *Sidereus Nuncius* Galileo referred to a second but different body of optical knowledge. Galileo explained that he improved the telescope on the basis of his knowledge of the *science of refraction*.[58] As we showed elsewhere, Galileo in fact could have calculated the properties of his superior lenses using Della Porta's theory of refraction.[59] This theory of refraction develops the essential quantitative geometrical relations by which the properties of optical elements are determined: (1) The relation between the radius of curvature and the angle of reflection/refraction of the incident ray; and, (2) The relation between the radius of curvature and the point at which the reflected/refracted ray intersects with the optical axis. These are elements of a theory by which specifications of lenses can be calculated and produced accordingly.[60]

---

[55] Note that in *Il Saggiatore* of 1619, Galileo publicly stated, for the first time, that the "telescope magnifies distant objects by presenting them to the eye in a larger angle than the angle under which they are seen without the instrument." Favaro (1890–1909, 6: p. 254): "Il telescopio rappresenta gli oggetti maggiori, perchè gli porta sotto maggiore angolo che quando son veduti senze lo strumento."

[56] Favaro (1890–1909, 10: p. 250, letter 228): "con un nuovo artifizio di un occhiale cavato dalle più recondite speculazioni di prospettiva"; Favaro (1890–1909, 10: p. 253, letter 231): "e parendomi che dovessi havere fondamento su la scientia di prospettiva, mi messi a pensare sopra la sua fabbrica." For an account on how the scientific community comprehended *perspective* and the working of visual aids then, at those days, see Smith A. M. (2001), Lindberg and Cantor (1985), Brownson (1981), Smith A. M. (1981), Kemp (1978), Lindberg (1976), Ronchi (1967; 1963).

[57] Zik and Hon (2012, pp. 442–445, 456).

[58] Galileo ([1610] 1989, p. 37), Favaro (1890–1909, 3: p. 60): "doctrinae de refractionibus innixus."

[59] Zik and Hon (2012, pp. 454–458).

[60] Zik and Hon (2012, pp. 459–460).

At the outset of *Sidereus Nuncius* Galileo stipulated the properties a good astronomical telescope should have. He emphasized that unless these properties are obtained, "one will try in vain to see all the things observed in the heavens by us and enumerated below [in *Sidereus Nuncius*]."[61] To convince his readers of the reliability of his observations Galileo described several procedures which reveal a few features of the telescope and how its optical performance can be calibrated. Galileo signaled how the optical system he constructed can realize its set of objectives operationally. This knowledge is essential for reproducing the observations which Galileo reported in *Sidereus Nuncius*.

What Galileo did was alien to the perspectivist qualitative theories of visual perception and the functioning of mirrors and lenses at the time. Galileo's manipulation of the focus shift, needed for adjusting the telescope (sect. 5.1) attests to his knowledge of the optical power of lenses which he linked to their "lengths". Galileo developed an original procedure for testing and calibrating his telescopes against a resolution target (sect. 5.2). He was able (1) to calculate the magnification of his telescopes, and (2) to modify the apertures so that only the right amount of light would enter the optical system. He realized that this is critical for minimizing disturbances and maximizing distinctness of the image. Galileo developed effective means for controlling most of the aberrations in telescopes composed of singlet lenses (sect. 5.3). He defined and calculated the apparent field of view of his telescopes and applied this optical knowledge for measuring angular distances between celestial bodies (sect. 5.4). He developed an improved micrometric device for measuring more accurately angular distances (sect. 5.5). Evidently, Galileo had knowledge of the relation between the magnitude of the image (in the image space) and that of the object (in the object space). Could these great achievements be accomplished by extending Ausonian optics of the spectacle makers in a lightly theorized practice based on "systematized experience"? We think not. We submit that Galileo had a novel optical theory which he did not want to divulge, but his practice and the extant instruments display it.

We give Kepler the final word. In April 19, 1610, in his response to Galileo's *Sidereus Nuncius*, Kepler remarked:

> So much for the instrument [telescope]. Now so far as its use is concerned, you [Galileo] have certainly discovered an ingenious method of ascertaining to what extant objects are magnified by your instrument, and how individual minutes and fraction of minutes can be discovered in the heavens.... Your achievements along these lines vies with Tycho Brahe's highly precise accuracy of observation....[62]

There is no better judge than Kepler to appreciate what Galileo had accomplished.

---

[61] Galileo ([1610] 1989, p. 38).
[62] Kepler ([1610] 1965, p. 21).

**Acknowledgments**   We thank Dov Freiman and A. Mark Smith for their valuable comments on earlier versions of this paper. This research is supported by the Israel Science Foundation (grant no. 67/09).

**References**

Baxandall, David. 1924. Replicas of Two Galileo Telescopes. *Transactions of the Optical Society* 25: 141–144.

Biagioli, Mario. 2006. *Galileo's Instruments of Credit: Telescopes, Images, Secrecy*. Chicago: The University of Chicago Press.

Biringuccio, Vannoccio. [1540] 1943. *The Pirotechnia of Vannoccio Biringuccio.* (trans: Smith, S. Cyril, and Gundi, Martha). New York: The American Institute of Mining and Metallurgical Engineers.

Bernieri, Eneico. 2012. Learning From Galileo's Errors. *Journal of the British Astronomical Association* 122: 169–172.

Borellio, Alphonso. 1666. *Theoricae Mediceorvm Planetarum*. Florentiae: Ex Typographia S.M.D.

Brown, Harold. 1985. Galileo on the Telescope and The Eye. *Journal of the History of Ideas* 46: 487–501.

Brownson, C. D. 1981. Euclid's optics and its compatibility with linear perspective. *Archive for history of Exact Sciences* 24: 165–194.

Della Porta, Giovan Battista. 1593. *De Refractione Optices Parte*. Naples: Apud Io. Iacobum Carlinum and Antonium Pacem.

Drake, Stillman. 1983. *Telescopes, Tides, and Tactics.* Chicago: The University of Chicago Press.

Drake, Stillman, and Kowal, Charles. 1980. Galileo's Sighting of Neptune. *Scientific American* 243 (6): 52–60.

Dupré, Sven. 2005. Ausonio's Mirrors and Galileo's lenses: The Telescope and Sixteenth Century Practical Optical Knowledge. *Galilaeana* 2: 145–180.

Favaro, Antonio. (ed.). 1890–1909. *Le Opere di Galileo Galilei*. Edizione Nazionale, 21 vols. Florence: G. Barbera, reprinted 1929–1939, 1964–1966.

Fuchs, E. Hofrat. 1901. *Text Book of Ophthalmology*. New York: D. Appleton and Company.

Galileo, Galilei. [1610] 1989. *Sidereus Nuncius or the Sidereal Messenger* (trans: Van Helden, Albert). Chicago: The University of Chicago Press.

Garzoni, Thomaso. 1587. *La Piazza Vniversale Di Tvtte Le Professioni Del Mondo*. Venetia: Appreso Gio. Battista Somasco.

Gingerich, Owen, and Van Helden, Albert. 2003. From *Occhiale* to Printed Page: The Making of Galileo's Sidereus Nuncius. *Journal for the History of Astronomy* 34: 251–267.

Greco, Vincenzo, et al. 1992. Optical Tests of Galileo's Lenses. *Nature* 358: 101.

Greco, Vincenzo, et al. 1993. Telescopes of Galileo. *Applied Optics* 32: 6219–6226.

Goldstein, R. Bernard, and Hon, Giora. 2005. Kepler's Move from Orbs to Orbits: Documenting a Revolutionary Scientific Concept. *Perspectives on Science* 13: 74–111.

Hofstetter, Henry. 1988. Optometry of Daza de Valdés. *American Journal of Optometry and Pysiological Optics* 65 (5): 354–357.

Ilardi, Vincent. 2007. *Renaissance Vision from Spectacles to Telescope*. Philadelphia: American Philosophical Society.

Kemp, Martin. 1978. Science non science and nonsense: The interpretation of Brunelleschi's perspective. *Art History* 1: 134–161.

Kepler, Johannes. [1610] 1965. *Conversation With Galileo's Sidereal Messenger* (trans: Rosen, Edward). London: Johonson Reprint Corp.

Lattis, James. 1994. *Between Copernicus and Galileo: Christoph Clavius and the Collapse of Ptolemaic Cosmology*. Chicago: University of Chicago Press.

Lindberg, David. 1976. *Theories of Vision from Al-Kindi to Kepler*. Chicago: University of Chicago Press.

Lindberg, David, and Cantor, Geoffry. 1985. *The Discourse of Light from the Middle Ages to the Enlightenment*. Los Angeles: University of California.

Malacara, Daniel. (ed.).(1992. *Optical Shop Testing*, 2nd edition. New York: Wiley.

Molesini, Giuseppe, and Greco, Vincenzo. 1996. Galileo Galilei: Research and Development of the Telescope, 423–438. In *Trends in Optics: Research, Developments and Applications*, (ed: Consortini, Anna). St Louis: Academic Press.

Pettit, Edison. 1939. A Telescope of Galileo. *Astronomical Society of the Pacific* 51: 147–150.

Ronchi, Vasco. 1923. Sopra I cannocchiali di Galileo. *L'Universo* 4: 791–804.

Ronchi, Vasco. 1963. Complexities, advances, and misconceptions in the development of science of vision: What is being discovered, 542–561. In *Scientific Change*, (ed: Alister Crombie). London: Heinemann.

Ronchi,V. (1967). The influence of the early development of optics on science and philosophy, 195–206. In *Galileo Man of Science*, (ed: Mcmullin, Ernan). New York: Basic Books.

Smith, A. Mark. 1981. Saving the Appearances of the Appearances: the Foundation of Classical Geometrical Optics. *Archive for History of Exact Sciences* 24: 73–99.

Smith, A. Mark. 2001. Practice vs. Theory: the Background to Galileo's Telescope Work. *Atti della Fondazione Giorgio Ronchi* 1: 149–162.

Smith, A. Mark. 2010. Alhacen on Refraction: A Critical Edition, with English Translation and Commentary, of Book 7 of Alhacen's *De Aspectibus*. *Transaction of the American Philosophical Society,* Volume one.

Smith, Warren. 1990. *Modern Optical Engineering*, 2nd edition. New York: McGraw-Hill.

Strano, Giorgio. 2009. Galileo's Telescope: History, Scientific Analysis, and Replicated Observations. *Experimental Astronomy* 25: 17–31.

Van Helden, Albert. 1977. The Invention of the Telescope. *Transaction of the American Philosophical Society* 67: 3–67.

Van Helden, Albert. (1999). *Catalogue of Early Telescopes.* Florence: Institute and Museum of the History of Science.

Van Helden, Albert. 2009. The beginnings, from Lipperhey to Huygens and Cassini. *Experimental Astronomy* 25: 3–16.

Van Helden, Albert. 2010. Galileo and the Telescope, 183–203. In Van Helden, Albert, et al. (2010).

Van Helden, Albert, et al (eds.). 2010. *The Origins of the Telescope*. Amsterdam: KNAW Press.
Willach, Rolf. 2008. *The Long Route to the Invention of the Telescope*. Philadelphia: American Philosophical Society.
Zik, Yaakov. 1999. Galileo and the Telescope: The Status of Theoretical and Practical Knowledge and Techniques of Measurement and Experimentation in the Development of the Instrument. *Nuncius* 14: 31–67.
Zik, Yaakov. 2001. Science and Instruments: The Telescope as a Scientific Instrument at the Beginning of the Seventeenth Century. *Perspectives on Science* 9: 259–284.
Zik, Yaakov. 2002. Galileo and Optical Aberrations. *Nuncius* 12: 455–465.
Zik, Yaakov, and Hon, Giora. 2012. Magnification: How to turn a spyglass into an astronomical telescope. *Archive for History of Exact Sciences* 66: 439–464.
Zik, Yaakov, and Van Helden, Albert. 2003. Between Discovery and Disclosure: Galileo and the Telescope, 173–190. In Bereta, Marco, et al. (eds.), *Musa Musaei*. Firenze: Leo S. Olschki.